\documentclass[aps,
	prl,
	superscriptaddress,
	reprint,
	twocolumn,
	amsfonts,
	amssymb,
	amsmath,
]{revtex4-1}

\usepackage{graphicx}
\usepackage{color}

\newcommand{\sign}{\text{sign}}
\newcommand{\R}{\mathcal{R}}

\newcommand{\Votk}{V_{\kk12}}
\newcommand{\Rotk}{\R_{12}^{\kk}}
\newcommand{\Rkto}{\R_{\kk2}^{1}}
\newcommand{\Rkot}{\R_{\kk1}^{2}}
\newcommand{\Rotkh}{\R_{12}^{\kk(h)}}
\newcommand{\Rktoh}{\R_{\kk2}^{1(h)}}
\newcommand{\Rkoth}{\R_{\kk1}^{2(h)}}

\newcommand{\kk}{\boldsymbol{k}}

\newcommand{\eez}{\boldsymbol{e}_z}

\newcommand{\nk}{n_{\kk}}

\newcommand{\Stk}{St_{\kk}}
\newcommand{\fk}{\mathcal{F}_{\kk}}
\newcommand{\dk}{\mathcal{D}_{\kk}}

\newcommand{\diff}{\mathrm{d}}
\newcommand{\dt}{\diff t}

\newcommand{\tnl}{t_{{\rm nl},\kk}}

\newcommand{\ok}{\omega_{\kk}}
\newcommand{\xik}{\xi_{\kk}}

\newcommand{\oo}{\omega_{1}}

\newcommand{\ot}{\omega_{2}}

\newcommand{\amin}{a_{\rm min}}
\newcommand{\qmin}{q_{\rm min}}
\newcommand{\qmax}{q_{\rm max}}

\newcommand{\khmin}{k_{h{\rm ,min}}}
\newcommand{\kzmin}{k_{z{\rm ,min}}}

\newcommand{\khmax}{k_{h{\rm ,max}}}
\newcommand{\kzmax}{k_{z{\rm ,max}}}

\newcommand{\kdinf}{k_{\rm d,inf}}
\newcommand{\kdsup}{k_{\rm d,sup}}

\newcommand{\kfh}{k_{{\rm f}h}}
\newcommand{\kfz}{k_{{\rm f}z}}
\newcommand{\of}{\omega_{\rm f}}
\newcommand{\xif}{\xi_{\rm f}}

\begin{document}

\title{Wave-kinetic dynamics of forced-dissipated turbulent internal gravity waves}

\author{Vincent Labarre}
\email[]{vincent.labarre@oca.eu}
\affiliation{Universit\'{e} C\^{o}te d'Azur, Observatoire de la C\^{o}te d'Azur, CNRS, Laboratoire Lagrange, Nice, France.}

\author{Giorgio Krstulovic}
\email[]{giorgio.krstulovic@oca.eu}
\affiliation{Universit\'{e} C\^{o}te d'Azur, Observatoire de la C\^{o}te d'Azur, CNRS, Laboratoire Lagrange, Nice, France.}

\author{Sergey Nazarenko}
\email[]{sergey.nazarenko@unice.fr}
\affiliation{Universit\'{e} C\^{o}te d'Azur, CNRS, Institut de Physique de Nice - INPHYNI, Nice, France}

\begin{abstract}
	Internal gravity waves are an essential feature of stratified media, such as oceans and atmospheres. To investigate their dynamics, we perform simulations of the forced-dissipated kinetic equation describing the evolution of the energy spectrum of weakly nonlinear internal gravity waves. During the early evolution, three well-known nonlocal interactions, the Elastic Scattering, the Induced-Diffusion, and the Parametric Subharmonic Instability, together with a Superharmonic Resonance play a prominent role. In contrast, local interactions are responsible for anisotropic energy cascade on longer time scales. We reveal emergence of a condensate at small horizontal wavevectors that can be interpreted as a pure wave-wave interaction-mediated layering process. 
\end{abstract}

\maketitle

When  a stable stratification gradient of salinity or temperature is present in a fluid, a peculiar type of wave can propagate in its bulk \cite{vallis_atmospheric_2017}. Such perturbations are called internal gravity waves (IGWs), and unlike most common waves, their wavelenghts and wave frequencies are not monotonically related; the IGW dispersion relation is given by $\ok=N k_h/|\kk|$, where $\kk$ is the wavevector, $k_h$ its horizontal magnitude (in the plane perpendicular to gravity) and $N$ the Brunt-Väisälä frequency \cite{Vallis_2017}. The inherently anisotropic character of the IGWs makes their dynamics rich and complex, even at the linear level \cite{maasObservationInternalWave1997,dauxoisInstabilitiesInternalGravity2018}. The IGWs also play a crucial role in ocean mixing \cite{muller_nonlinear_1986, lvov_resonant_2012,mackinnon_climate_2017, dematteis_interacting_2024}.
One of the first descriptions of the ocean internal wave field was given by Garrett and Munk (GM) \cite{garrett_space_1972, garrett_space_1975}, who furnished an empirical expression for the wave energy spectra. Since then, many theoretical works have followed GM phenomenology to explain the mechanisms behind the formation of such spectra. 
From a more theoretical side, the Weak Wave Turbulence  (WWT) theory \cite{peierls_kinetischen_1929, hasselmann_Feynman_1966, zakharov_kolmogorov_1992, nazarenko_wave_2011} provides a wave kinetic equation (WKE) describing the evolution of IGWs under the effect of weakly nonlinear interactions. This equation has been derived using various approaches and approximations; see, e.g., \cite{olbers_nonlinear_1976,caillol_kinetic_2000,lvov_hamiltonian_2001,lvov_hamiltonian_2004,Labarre_KineticsInternalGravity_2024}. It always takes the form $\dot{n}_{\kk} = \Stk[\nk]$, where $\kk$ is the wave-vector, and $\nk$ is the 3D wave action spectrum, which is proportional to the magnitude square of the Fourier transform of the wave amplitude. $\Stk$ is the collision integral describing interactions between waves. The WWT is a general theory that not only applies to IGWs, but to many other systems such as plasma waves \cite{Zakharov_CollapseLangmuirWaves_}, gravitational waves \cite{galtier2017turbulence}, turbulence in Bose-Einstein condensates \cite{Dyachenko:1992aa,zhuDirectInverseCascades2023}, and so on \cite{nazarenko_wave_2011}. The theory typically furnishes analytical predictions that can be tested in experiments and numerical simulations.

\begin{figure}
	\centering
	\includegraphics[width=\columnwidth]{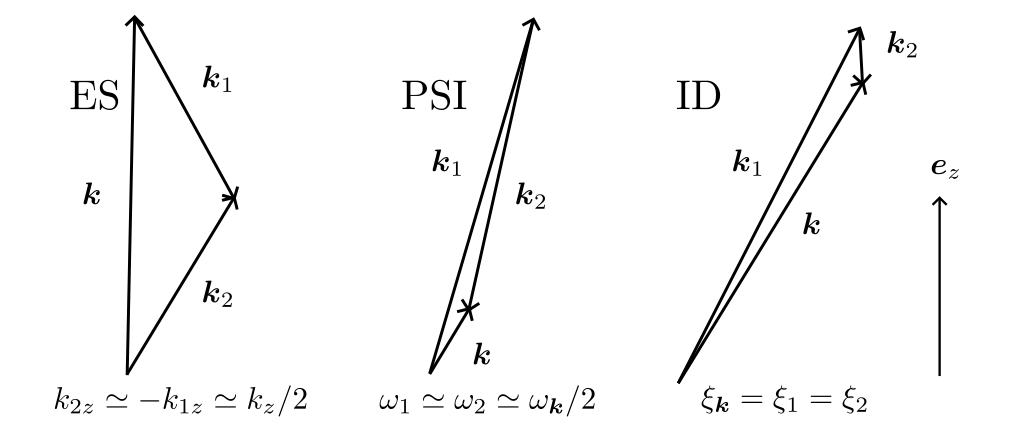}
	\caption{A cartoon of the three nonlocal interactions identified by \citet{mccomas_bretherton_resonant_1977}. $\eez$ is the vertical axis. \label{fig:NonLocalInterac}}
\end{figure}

The WKE gives access to the wave dynamics on kinetic (very large) time scales that cannot be easily accessed through direct numerical simulations of the fluid dynamical equation in the weakly nonlinear regime. In many fields, numerical simulations of WKEs have become a powerful numerical tool to describe nonlinear wave cascades and transient states \cite{Galtier_WeakTurbulenceTheory_2000a,Pan_Yue_2017,zhuDirectInverseCascades2023,zhuSelfsimilarEvolutionWave2023,eremin2024wave}. For example, WKE simulations are central to the so-called ``wave models'' used for sea state forecasts \cite{tolman1991third}.
In the case of IGWs, simulations of the WKE are delicate and have a cost $\propto (\# {\rm grid \,points})^4$, which has challenged the study of steady-state turbulent IGW. 
For this reason, early numerical IGW works in the 70's focused on the estimation of energy transfers by computing the IGW collision integral for some test spectra (particularly for GM) \cite{olbers_nonlinear_1976,mccomas_bretherton_resonant_1977, mccomas_time_1981, mccomas_dynamic_1981, muller_nonlinear_1986}.
Those studies revealed that nonlocal interactions (i.e., involving modes with very different wave-vector lengths or wave frequencies) are essential. Namely, three mechanisms were identified by \citet{mccomas_bretherton_resonant_1977}:
\begin{itemize}
	\item[ES:] \emph{Elastic Scattering}. Two waves of similar vertical wave number modulus $k_{2z}\backsimeq -k_{1z}$, interact with another one having a lower frequency and twice the vertical wavenumber $k_z=k_{2z}-k_{1z} \backsimeq 2 k_{2z}$. 
	\item[PSI:] \emph{Parametric Subharmonic Instability}. A low-wavenumber mode decays into two with high wavenumbers with half the frequency, i.e. $\ok/2\backsimeq\oo\backsimeq\ot$.
	\item[ID:] \emph{Induced Diffusion}. One low-frequency wave interacts with two approximately identical waves of much larger wave number and frequency. As shown in \cite{lanchon_energy_2023}, ID proceeds along lines $\xik\equiv k_h/k_z^2=k_{1h}/k_{1z}^2=k_{2h}/k_{2z}^2=const$, similar to the case of nonlocal drift waves \cite{balk_nonlocal_1990}.
\end{itemize}
A sketch of the three nonlocal interactions is displayed in Fig.~\ref{fig:NonLocalInterac}. Moreover, a fourth nonlocal interaction is also relevant \cite{sun2012energy,husseini_experimental_2020}:
\begin{itemize}
	\item[SR:] \emph{Superharmonic Resonance}. Two waves having similar frequencies $\oo \backsimeq \ot$ generate a third wave with twice frequency $\ok \backsimeq 2 \oo$.
\end{itemize}
More recent studies revisited the energy transfers for specific spectra, and showed that both, local and nonlocal interactions are important \cite{dematteis_origins_2022, wu_energy_2023, taebel2024experimental}. Therefore, all interactions should be considered when dealing with the IGW evolution, which makes the analytical description of such a system difficult. 

The WWT predicts non-equilibrium, scale-invariant steady-state spectra \cite{pelinovsky_raevsky_weak_1977, caillol_kinetic_2000, lvov_hamiltonian_2001} that, in the hydrostatic limit, are bi-homogeneous in horizontal and vertical wave-vector amplitudes $k_h$ and $|k_z|$ (i.e., spectra $\propto k_h^{\nu_h} |k_z|^{\nu_z}$). 
However, a rigorous analysis showed that most of these spectra are not mathematically valid solutions because the collisional integral diverges \cite{lvov_oceanic_2010}. The only exception is a spectrum found numerically on the line $\nu_z=0$, which corresponds to the exact cancelation of two divergent parts of the collision integral \cite{lvov_oceanic_2010, dematteis_downscale_2021}. Note that for this solution, any perturbation can lead to a divergence of the collision integral, which puts the physical realizability of this spectrum in question. Finally, considering only nonlocal ID interactions yields yet another analytical prediction \cite{lanchon_energy_2023}.

Current theoretical predictions are all limited to steady-state spectra and are hardly realizable in experiments or direct numerical simulations of the Navier-Stokes-Boussinesq equations. In this Letter, we present numerical simulations of the forced-dissipated WKE for IGWs in the hydrostatic limit, i.e. $k_h\ll|k_z|$. The hydrostatic approximation, commonly used in oceanographic theoretical studies \cite{lvov_oceanic_2010,dematteis_origins_2022,dematteis_interacting_2024}, allows for great simplifications as the resonant manifold (defined later) can be treated analytically and the WKE becomes bi-homogenous in $k_h$ and $|k_z|$. The latter property allows us to place a localized forcing in any sector of our computational box to explore different regimes. Our simulations allow us to observe the evolution of the energy spectrum due to weakly nonlinear wave-wave interactions and the establishment of a nontrivial steady-state spectrum.

We consider the WKE derived by \citet{lvov_hamiltonian_2001}, to which we add a forcing and a dissipative term:
\begin{align}
	\dot{n}_{\kk} &= St_{\kk} + \fk - \dk \nk \label{eq:WKE}\\
	St_{\kk} &= \int ~ \left[ \Rotk - \Rkto - \Rkot \right] ~ \diff^3 \kk_1 \diff^3 \kk_2 \\
	\Rotk &= 4 \pi ~ \delta(\kk-\kk_1-\kk_2) ~ \delta(\ok-\oo-\ot) \\
	\nonumber
	& \quad \times ~ |\Votk|^2 ~ \left( n_1 n_2 - n_{\kk} n_1 - n_{\kk} n_2 \right)
\end{align}
where $\ok = N k_h /|k_z|$ is the hydrostatic wave frequency and $\Votk$ is the interaction coefficient; see the supplemental material (SM) for details. Waves  interact only if their corresponding wavevectors lay on the resonant manifold defined by the equations
\begin{eqnarray}
	\kk = \kk_1 + \kk_2 &,&\ok = \oo + \ot\label{eq:resManifold},
\end{eqnarray}
and the respective equation with cyclic permutation of indexes. 

We consider axisymmetric and vertically symmetric spectra, i.e., depending only on $(k_h,|k_z|)$, such that the 2D wave energy spectrum is defined as
\begin{equation}
	e(k_h,k_z) = 4 \pi k_h \ok \nk.
\end{equation}
Note that with this definition, the total energy per unit of volume is $E=\int_{0}^\infty \int_{0}^\infty e(k_h,k_z)\diff k_h \diff k_z$. The dissipation coefficient $\dk$ in \eqref{eq:WKE} is defined such that energy is efficiently dissipated for $k_h, |k_z| \lesssim \kdinf$ and $\sqrt{k_h^2 + k_z^2} \gtrsim \kdsup$. In this Letter, we consider a localized forcing $\fk$ centered at $(\kfh,\kfz)$, so that the rate of the energy injection is $P=4 \pi \int_{0}^\infty \int_{0}^\infty \ok \fk k_h\diff k_h \diff k_z$. We express time in units of the nonlinear time scale $\tau_{\rm nl}=(k_f^2 P/\of)^{-1/2}$, with $1/k_f$ being the typical forcing length scale and $\of=N\kfh/\kfz$ is the typical frequency of the forced waves.
We use a logarithmic grid of $M_h \times M_z$ points spanning the spectral domain $[\khmin;\khmax] \times [\kzmin;\kzmax]$. We fix the dissipative scales $\kdinf$ and $\kdsup$ close to the boundaries of the $k$-domain to enlarge the inertial range but ensuring a well resolved spectrum. For each grid point, the collision integral is given by a 2D integral, obtained after the parametrization of the resonant manifold and which is computed using the trapezoidal rule on the log-grid. The numerical cost per time-step is then of the order of $M_h^3 M_z$ operations. The collisional integral has integrable divergence that are treated analytically. Intermesh values are obtained using a bi-linear interpolation scheme, which preserves positivity. Time advancement is done by the Runge-Kutta 2 method, with an adaptive time step based on the time scale of the collision integral. See the SM for more details on the WKE, its resonant manifold and on the numerical method. In this Letter, we present a short simulation spanning a large spectral domain, and a longer one on a reduced domain using $M_h=M_z=80$ grid points in both simulations. Moreover, to emphasize the robustness of our results, the SM inlcudes some simulations with different types of forcings and initial conditions. 

\begin{figure}
	\centering
	\includegraphics[width=\columnwidth]{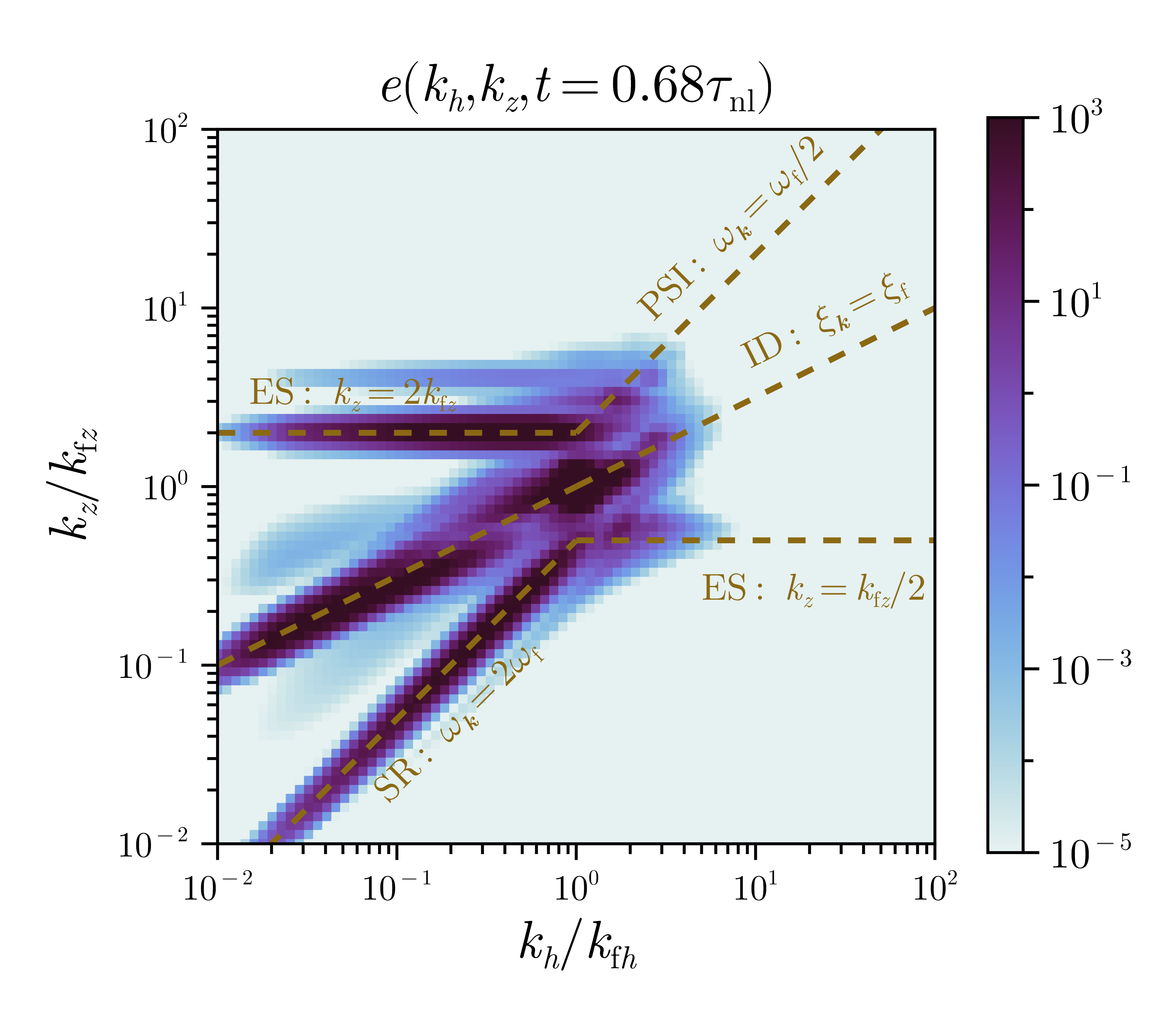}
	\caption{Energy spectrum $e(k_h,k_z,t)$ in the transient state for a simulation with a localized forcing in the middle of the $(k_h,|k_z|)$-space. The dashed lines indicate wave vectors corresponding to nonlocal interactions with the forced wave vectors $(\kfh,|\kfz|)$: $k_z=2\kfz$ and $k_z=\kfz/2$ (ES), $\ok=\of/2$ (PSI), $\xik=\xif$ (ID), and $\ok=2\of$ (SR). \label{fig:spectrumFull}}
\end{figure}

We first run a simulation with forcing located in the middle of the $(k_h,k_z)$-space, which spans four decades in each direction. Initially, the energy spectrum is zero everywhere, and then it quickly develops along three main lines in the $k_h,k_z<k_{\rm f}$ sector, as shown in Fig.~\ref{fig:spectrumFull} at $t=0.68\tau_{\rm nl}$. Evolution's timescale is about a nonlinear time.
The horizontal lines correspond to the ES, with $k_{2z}=-k_{1z}=\kfz$ so that $k_z=2\kfz$. ID proceeds along the line $\xi=\xif=\kfh/\kfz^2$ as shown in \cite{lanchon_energy_2023}, similarly to drift waves \cite{balk_nonlocal_1990}. Finally, the third line corresponds to the SR with the forcing at $\ok=2\of$. Note that these excited branches are responsible for an energy transfer towards large scales. Such an inverse energy transfer is consistent with the fact that IGWs are known to concentrate energy at large horizontal scales \cite{billant_self_2001, smith_generation_2002, brethouwer_scaling_2007,remmel_nonlinear_2014,Labarre_InternalGravityWaves_2024}. Here, such an inverse energy transfer is the result of the resonant three-wave  interactions only, as observed in \cite{Scott_evolution_2024}. We also notice the development of the spectrum along three other lines, toward small scales $k_h,k_z<k_{\rm f}$. The first line is $k_z=\kfz/2$ and is explained by the ES triadic interactions \cite{wu_energy_2023}. The second line corresponds to the PSI with the forcing at $\ok=\of/2$. The third line is due to the ID.

\begin{figure}
	\centering
	\includegraphics[width=\columnwidth]{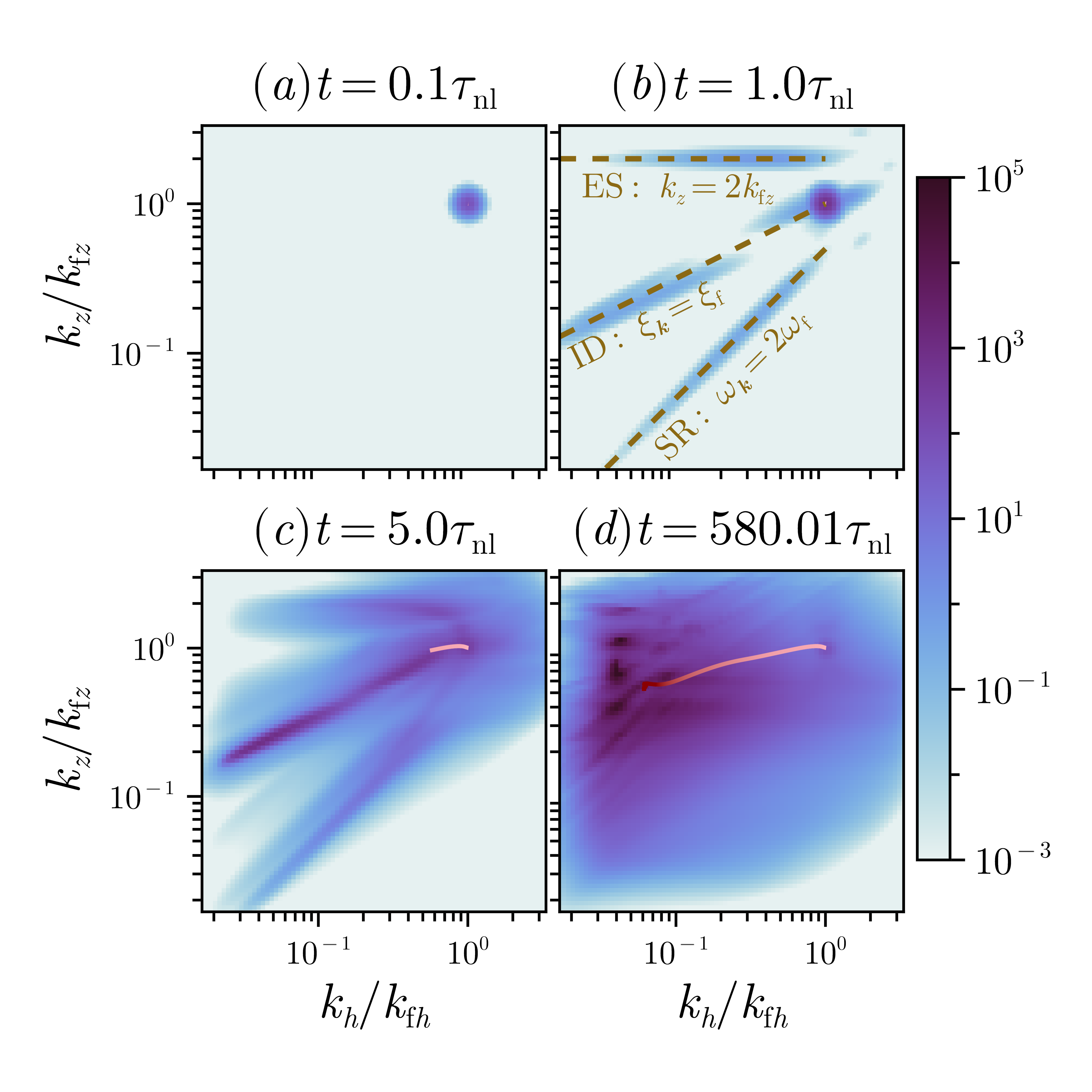}
	\caption{Snapshots of the energy spectrum $e(k_h,k_z,t)$ at different times for the simulation with a small-scale forcing. On the panel (b), dashed lines are for $k_z=2\kfz$, $\ok=2\of$ and $\xik=\xif$, corresponding respectively to ES, SR, and ID. The colored line represents the trajectory $(K_h(t),K_z(t))$, defined in equation (\ref{eq:IntegralScale}), with dark red colors corresponding to larger $t$.
		\label{fig:spec2D_long_evol}}
\end{figure}

Propagation of the spectrum toward higher wavevectors, including the PSI, is much slower and computationally demanding. In this Letter, we will focus on the inverse energy transfers in the $(k_h,k_z)$-space, which have an important impact on the IGWs dynamics and have received less attention than the other type of transfers \footnote{In the SM, we investigate the effect of forcing shape and width, employing simulations at lower resolution.}. Consequently, we now set the forcing at high-wavevectors and run a new simulation in a smaller box until the system reaches a quasi-steady state. We show the evolution of the energy spectrum in Fig.\ref{fig:spec2D_long_evol}. The colored curves in panels (c) and (d) show the trajectory of the integral scales 
\begin{equation}
	\label{eq:IntegralScale}
	\left(\frac{1}{K_h(t)},\frac{1}{K_z(t)}\right)=\frac{1}{E}\int \left(\frac{1}{k_h},\,\frac{1}{k_z}\right)e(k_h,k_z) \diff k_h \diff k_z
\end{equation}
in temporal interval $[0,t]$.

Panel (a) displays a very early time, with almost all the energy being contained in the forced modes. Panel (b) shows the same instabilities as in Fig.~\ref{fig:spectrumFull}. Panel (c), taken at five nonlinear times, shows that energy starts spreading from the branches excited by the nonlocal interactions, in a process resembling to a local energy cascade, but completely anisotropic.
We also observe that the integral scale of the flow has moved away form the forcing region. Finally, panel (d) displays the energy spectrum at the end of our simulations, when the system is very close to a steady state. 
We observe that the energy can be transferred to higher frequencies (i.e. $k_h \gg k_z$), which is consistent with other studies \cite{dematteis_downscale_2021,wu_energy_2023}. Moreover, nonlocal interactions have transferred most of the energy toward large scales and small frequencies, and local interactions have smoothened out the spectrum so that the forcing region is hardly visible. Interestingly, a new peak emerges at low $k_h$ and high $k_z$ that overwhelms the energy spectrum, as manifested by the evolution of integral scale that has stopped close to this scale. Such a peak resembles a condensate and its formation could be interpreted as a layering process, which is often observed in experiments and simulations \cite{caulfield2021layering}. Layering can be seen as the condensation of energy at the lowest wave frequency for stratified flows. In the presence of rotation, the condensation will be saturated at the rotation frequency. It is similar to the formation of the near-inertial peak observed in the ocean. Remarkably, the emergence of layers is produced by nonlinear wave interactions only, as shear modes and the vertical vorticity are not taken into account in the wave-kinetic description of IGWs.

The state observed in Fig.~\ref{fig:spec2D_long_evol}.d corresponds to a non-equilibrium turbulent solution of the WKE \eqref{eq:WKE}, in which a localized energy input by the forcing balances the dissipation acting at well-separated scales. Turbulent solutions are often characterized by the emergence of power-law spectra in the inertial range, i.e. in between (and far from) the forcing and the dissipation scales. IGWs are highly anisotropic, but their WKE collision integral kernel is bi-homogenous. It is then natural to look for solutions of the form $\nk \propto k_h^{\nu_h} |k_z|^{\nu_z}$. Moreover, due to the properties of the ES, the PSI, and the ID, it is also natural to consider solutions in the $(\ok,k_z)$-plane, as it is typically done in oceanographic studies \cite{dematteis_interacting_2024}. A simple algebra, and a change of variable, gives for the energy spectrum
\begin{equation}
	e(k_h,k_z) \propto k_h^{\nu_h + 2}k_z^{\nu_z-1},{\rm and}\, e(\ok,k_z) \propto \ok^{\nu_h+2}k_z^{\nu_h+\nu_z +2}.
\end{equation}
Some values of $(\nu_h,\nu_z)$ can be found in the literature, which we discuss in the following. The high-frequency limit of the GM spectrum has the exponents $(-4,0)$ \cite{garrett_space_1972}, whereas the Rayleigh-Jeans thermal equilibrium spectrum corresponds to $(-1,1)$  \cite{caillol_kinetic_2000}. The steady state spectrum obtained numerical by \citet{lvov_oceanic_2010} corresponds to $(-3.69,0)$ and the nonlocal ID spectrum of \citet{lanchon_energy_2023} to  $(-3,-1)$.

\begin{figure}
	\centering
	\includegraphics[width=\columnwidth]{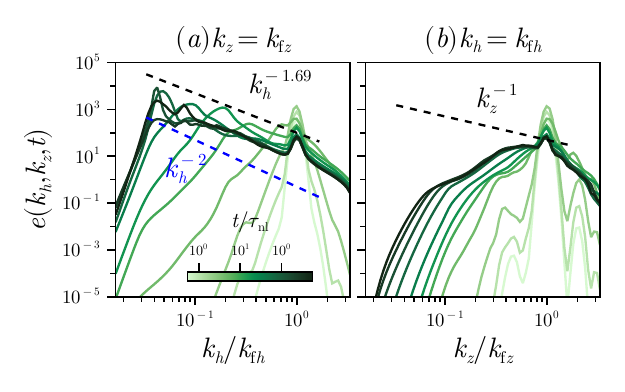}
	\caption{Slices of the energy spectrum at different times: (a) $e(k_h,k_z=\kfz,t)$ and (b) $e(k_h=\kfh,k_z,t)$. }\label{fig:cuts_kzkh}
\end{figure}

Figures \ref{fig:cuts_kzkh} (a) and (b), show slices of the 2D energy spectrum shown in Fig.\ref{fig:spec2D_long_evol}(d) for fixed values of, respectively, $k_h= \kfh$ and $k_z= \kfz$ at different times. 
In Fig.\ref{fig:cuts_kzkh}(a), we clearly observe the early stage where energy increases at the forcing scale, followed by energy transfer to larger scales, akin to an inverse cascade with a spectral front propagation. In the figure, we display the GM ($k_h^{-2}$) and the \citet{lvov_oceanic_2010} ($k_h^{-1.69}$) spectral slopes for comparison. Although both scalings seem to be in reasonable agreement with numerics, one has to be careful with such a comparison because the full 2D spectrum deviates from the bi-homogenous shape implied by both GM and \citet{lvov_oceanic_2010} spectra. At the latest times, a condensate peak is clearly visible at a small $k_h$. 
The evolution of the spectrum for a fixed $k_h$, shown in Fig.\ref{fig:cuts_kzkh}(b), is more intricate. At wavevectors larger than $\kfz$, a spectrum steeper than $k_z^{-1}$, expected for GM and the prediction of \cite{lvov_oceanic_2010}, is observed. 

\begin{figure}
	\centering
	\includegraphics[width=\columnwidth]{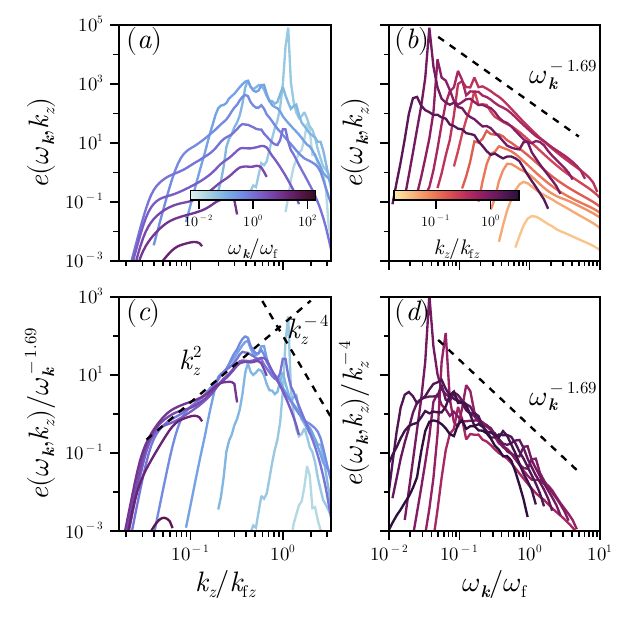}
	\caption{Slices of the spectrum $e(\ok,k_z)$ with fixed $\ok$ (panels (a) and (c)) and fixed $k_z$ (panels (b) and (d)). Panels (a) and (b) are uncompensated plots, and (c) and (d) are compensated  by $\ok^{-1.69}$ and $k_z^{-4}$ respectively. In panel (d), we show only lines corresponding to $k_z\gtrsim \kfz/2$. 
		\label{fig:cuts_kz_om}}
\end{figure}

Finally, in Fig.\ref{fig:cuts_kz_om} we display cuts of $e(\ok,k_z)$ at the latest time of our simulation.
Figures \ref{fig:cuts_kz_om}(a) and (b) show the raw data, whereas in (c) and (d) the spectra are compensated by the apparent scalings. When the spectrum is compensated by $\ok^{-1.69}$, a good collapse of the curves is obtained as shown in Fig.~\ref{fig:cuts_kz_om}.(c). The $\ok^{-1.69}$ scaling is also visible in Fig.~\ref{fig:cuts_kz_om}.(d), which has been compensated by the $k_z^{-4}$ scaling observed for $k_z>\kfz$. To our knowledge, such a steep $k_z$-dependence has not been reported before. However, it could result from the limited inertial range for $k_z>\kfz$. Finally, we remark that at wavevectors $k_z<\kfz$, the spectrum displays a scaling compatible with a thermal RJ $k_z^2$-spectrum. In summary, while the prediction $k_h^{-1.69}$ seems to work better than the famous GM spectrum, the agreement is only partial and is limited to some sectors and projections. This fact reflects the complexity of internal waves and perhaps the lack of pure bi-homogeneous power-law steady-state spectrum.

To compare our numerical simulations to oceans and experiments, note that in the WWT framework employed here,  $\tau_{\rm nl} = Fr^{-2} \tau_{\rm lin}$, where $\tau_{\rm lin} \sim N^{-1}$ is the characteristic linear time of waves and $Fr$ is the Froude number that compares the typical flow velocity to the restoring velocity of internal waves. Typical values in the ocean and laboratory experiments are $N=5\times 10^{-3}~rad.s^{-1}$ and $N\sim1~rad.s^{-1}$ respectively, whereas the Froude number is about $8.4\times 10^{-3}$ for the ocean \footnote{To compute the typical Froude in the ocean, we use the definition $Fr = U/(NL)$, and the typical values for GM spectrum (see \cite{lvov_resonant_2012}): a thermocline height $L=1300~m$, a typical velocity based on the amplitude of the GM spectrum $U = \sqrt{E_0} = \sqrt{30 \times 10^{-4}}~m.s^{-1}$ and the typical buoyancy frequency $N=5\times 10^{-3}~rad.s^{-1}$.} and $10^{-2}$ for experiments \cite{rodda_experimental_2022,Lanchon_InternalWaveTurbulence_2023}. That leads to a nonlinear time of roughly $\tau_{\rm nl}\sim 33$ days for the ocean and $\tau_{\rm nl} \sim 2.8$ hours for the experiments. Note that the first instabilities transfer energy towards the large scales in about $3\tau_{\rm nl}$ and the condensate peak emerges at about $100\tau_{\rm nl}$. Such time scales are consistent with values found in the literature \cite{mccomas_bretherton_resonant_1977,Lanchon_InternalWaveTurbulence_2023}.

In this Letter, we have reported numerical simulations of the wave-kinetic equation for weakly nonlinear IGWs. Whereas previous studies have focused on the evaluation of energy transfers for given spectra \cite{olbers_nonlinear_1976, eden_numerical_2019, olbers_psi_2020, dematteis_downscale_2021, dematteis_origins_2022, wu_energy_2023}, we have performed forced-dissipated simulations and looked at the spectrum evolution until it reaches a quasi-steady state. 

In our simulations, we have observed the evolution of the energy spectrum under the effect of both local and nonlocal interactions. We showed that for a localized forcing, the nonlocal interactions are initially faster, but the local interactions are important for smoothing out the spectrum on longer time scales. Energy transfers go in every direction of the spectral space: to small scales and small frequencies, but also to large scales and large frequencies. The energy transfer toward the large scales is initially quicker than the transfer toward the small scales. At the final state, a condensate peak emerges in the spectral space at a low value of $k_h$.
The observed dynamics has consequences for the formation of large-scale structures due to interacting waves, and for the small-scale diapycnal mixing. At the end of our simulation, the energy is less present at large frequencies, with a spectrum $e(k_h,k_z) \propto k_h^{-1.69}$ \cite{lvov_oceanic_2010,dematteis_downscale_2021} in some sectors. However, the $k_z$ slope is different for small $k_z$ and large $k_z$, and is incompatible with the Garrett-Munk phenomenology and theoretical predictions. It gives another support to the formation of slow modes under the effect of resonant wave-wave interactions.
Such an energy transfer toward the large scales could change the energy pathway to mixing and may have to be considered in fine scale parametrization of oceans. 

\begin{acknowledgments}
	We thank Nicolas Mordant, Giovanni Dematteis, Yue Cynthia Wu, Yulin Pan, Nicolas Lanchon, Pierre-Philippe Cortet, and Michal Shavit for their helpful discussions. We also gratefully acknowledge the valuable feedback from four anonymous referees, which significantly improved this work. Computations have been done on the ``Mesocentre SIGAMM'' machine, hosted by Observatoire de la Cote d'Azur. This work was supported by the Simons Foundation project "Collaboration in Wave Turbulence" (award ID 651471).
\end{acknowledgments}

\bibliography{biblio}

\clearpage
\appendix

\begin{widetext}
	\begin{center}
		\Large
		\textbf{Supplemental material of ``Wave-kinetic dynamics of forced-dissipated turbulent internal gravity waves''}
	\end{center}
	
	\renewcommand\thefigure{S\arabic{figure}}    
	\setcounter{figure}{0}
	
	\begin{center}
		This document gives technical details about the numerical resolution of the kinetic equation. Firstly, we describe the resonant manifold, the expression of the collision integral for axisymmetric and vertically symmetric spectra. Secondly, we show the numerical grids used for the computations. Thirdly, we present our integration and interpolation schemes. Finally, we explain our time-stepping scheme. The last sections provide additional studies based on wave kinetic equation simulations concerning the random perturbation of the initial condition and forcing shape and width.
	\end{center}

	\section{Collision integral and resonant manifold}
	
	For axisymmetric and vertically symmetric spectra, the kinetic equation can be written as \cite{dematteis_downscale_2021}
	\begin{align}
		\label{eq:KineticEquationAxisymmetric}
		\dot{n}_{\kk} &= \Stk[\nk] + \fk - \dk \nk = 8 \pi ~ \int ~ \left[ \Rotkh - \Rktoh - \Rkoth \right] ~ k_{1h} k_{2h} ~ \diff k_{1h} \diff k_{2h} + \fk - \dk n_{\kk} \\
		\Rotkh &= \frac{|\Votk|^2}{|g'| \Delta} ~ \left( n_1 n_2 - n_{\kk} n_1 - n_{\kk} n_2 \right) \\
		\Votk &= \sqrt{\frac{k_h k_{1h} k_{2h}}{32}} \left( \frac{\kk_h \cdot \kk_{1h}}{k_h k_{1h}} \sqrt{\left| \frac{k_{2z}}{k_z k_{1z}} \right|} + \frac{\kk_h \cdot \kk_{2h}}{k_h k_{2h}} \sqrt{\left| \frac{k_{1z}}{k_z k_{2z}} \right|} + \frac{\kk_{1h} \cdot \kk_{2h}}{k_{1h} k_{2h}} \sqrt{\left| \frac{k_z}{k_{1z} k_{2z}} \right|} \right) \\
		\Delta &= \frac{1}{2} \sqrt{(- k_h + k_{1h} + k_{2h}) (k_h - k_{1h} + k_{2h}) (k_h + k_{1h} - k_{2h}) (k_h + k_{1h} + k_{2h})} \\
		g' &= \frac{k_{1h} ~ \sign(k_{1z})}{k_{1z}^2} - \frac{k_{2h} ~ \sign(k_{2z})}{k_{2z}^2}
	\end{align}
	where $k_{1z}$, $k_{2z}$, and the scalar products of horizontal wave-vectors ($\kk_h \cdot \kk_{1h}$, $\kk_h \cdot \kk_{2h}$, and $\kk_{1h} \cdot \kk_{2h}$) are evaluated for the solution of the resonance conditions, i.e. $\kk = \kk_1 + \kk_2$ and $\ok = \oo + \ot$ (or permutations). Here, we have introduced a forcing term $\fk$ and dissipative term $\dk \nk$, with $\dk$ a dissipation coefficient. We fix
	\begin{equation}
		\dk = \frac{1}{\ok} \left[ \left(\frac{k}{\kdsup}\right)^{8} + \left(\frac{k_h}{\kdinf}\right)^{-8} + \left(\frac{|k_z|}{\kdinf}\right)^{-8} \right]  
	\end{equation}
	and 
	\begin{equation}
		\fk = \frac{f_0}{\ok} \exp \left[ - \frac{(\ln k_h - \ln \kfh)^2 + (\ln |k_z| - \ln |\kfz|)^2}{(\ln \Delta k_{\rm f})^2} \right],
	\end{equation}
	were $(\kfh,|\kfz|)$ are the wave vectors component amplitudes of the forced modes, $\Delta k_{\rm f}$ fixes the width of the forcing, and the normalization factor $f_0$ is fixed such that the energy injection rate (computed numerically) is equal to unity. \\
	
	The parametrization of the resonant manifold can be found in \citet{lvov_oceanic_2010}. For $\Rotkh$ $\kk = \kk_1 + \kk_2$, $\ok = \oo + \ot$, we have the two branches
	\begin{align}
		\label{eq:k1zkp}
		k_{1z} &= \frac{k_z}{2 k_h} \left( k_h + k_{1h} + k_{2h} + \sqrt{ (k_h + k_{1h} + k_{2h})^2 - 4 k_h k_{1h}} \right)  \\ 
		\label{eq:k1zkm}
		k_{1z} &= \frac{k_z}{2 k_h} \left( k_h - k_{1h} - k_{2h} - \sqrt{(k_h - k_{1h} - k_{2h})^2 + 4 k_h k_{1h}} \right) 
	\end{align}
	For $\Rktoh$ $\kk_1 = \kk + \kk_2$, $\oo = \ok + \ot$, we have the two branches
	\begin{align}
		\label{eq:k1z1p}
		k_{1z} &= \frac{k_z}{2 k_h} \left( k_h + k_{1h} + k_{2h} - \sqrt{(k_h + k_{1h} + k_{2h})^2 - 4 k_h k_{1h}} \right) \\ 
		\label{eq:k1z1m}
		k_{1z} &= \frac{k_z}{2 k_h} \left( k_h - k_{1h} + k_{2h} - \sqrt{(-k_h + k_{1h} - k_{2h})^2 + 4 k_h k_{1h}} \right) 
	\end{align}
	The $\Rkoth$ and $\Rktoh$ terms give the same contribution to the collisional integral, which is used to reduce the computational cost. Because $(\kk_{h}, \kk_{1h}, \kk_{2h})$ form a triad, they must satisfy the triangular inequalities
	\begin{equation}
		\label{eq:KinematicBox}
		k_h \leq k_{1h} + k_{2h}, ~~~~ k_{1h} \leq k_h + k_{2h}, ~~~~ k_{2h} \leq k_{1h} + k_h,
	\end{equation}
	meaning that $(k_{1h},k_{2h})$ must lie in the so called ``kinematic box'' shown in Fig.\ref{figure1}(a). It is therefore more convenient to work with the $(p,q)$ variables such that
	\begin{equation}
		\label{eq:PQ}
		k_{1h} = \frac{k_h + p + q}{2} ~~~~ \text{and} ~~~~ k_{2h} = \frac{k_h - p + q}{2}.
	\end{equation}
	The kinematic box is then given by the domain $(p,q) \in [-k_h,k_h] \times [0,\infty[$ and the collision integral reads
	\begin{equation}
		\Stk = 4\pi ~ \int ~ \left[ \Rotkh - 2 \Rktoh \right] ~ k_{1h} k_{2h} ~ \diff p \diff q 
	\end{equation}
	and
	\begin{equation}
		\Delta(p,q) = \frac{1}{2}\sqrt{(k_h^2 - p^2) q (2 k_h + q)}.
	\end{equation}
	
	\begin{figure}[h]
		\centering
		\includegraphics[width=\linewidth]{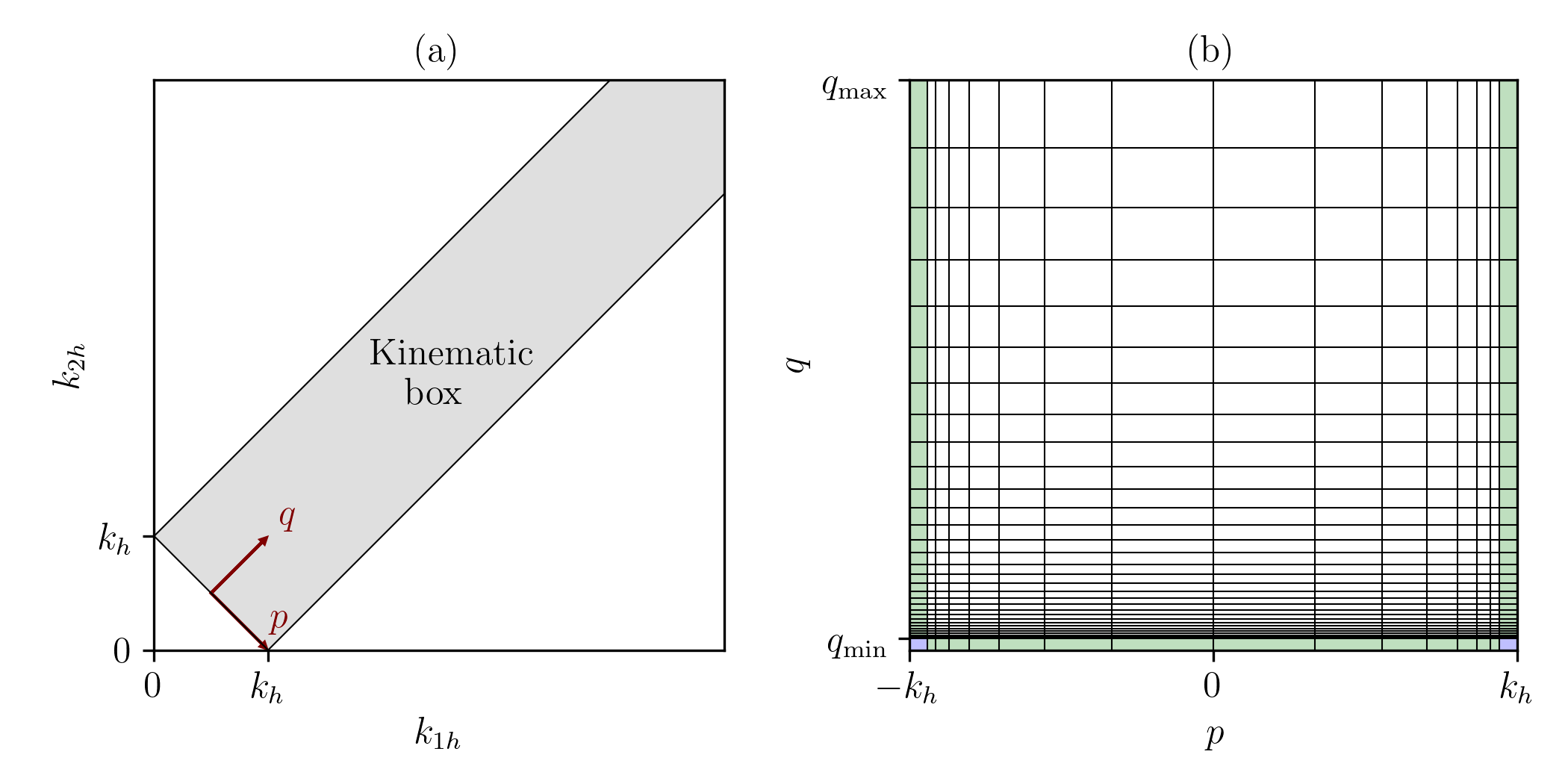}
		\caption{(a) Kinematic box defined by conditions (\ref{eq:KinematicBox}) and the $(p,q)$ coordinates (\ref{eq:PQ}). (b) Illustration of the grid used for the computation of the collision integral at a given $(k_h,k_z)$. Blue and green cells corresponds to regions where special treatment is required because of singularities (zeros of $\Delta(p,q)$). The size of these cells have been increased for visualization purposes. \label{figure1}}
	\end{figure}

	\section{Numerical grids}
	
	We store the spectra on a logarithmic grid of $M_h \times M_z$ points. Namely $k_\alpha$ takes values in 
	\begin{equation}
		\label{eq:LogGrid}
		k_\alpha[i] = k_{\alpha {\rm ,max}}\lambda_h^{-M_\alpha+i}, ~~~~ i \in [1:M_\alpha],
	\end{equation}
	with $\lambda_\alpha = (k_{\alpha {\rm,min}}/k_{\alpha {\rm,max}})^{1/(1-M_\alpha)}$ for $\alpha = h$ or $z$. For each $(k_h,k_z)$ of the grid, we compute $\Stk$ by integration over $(p,q) \in [-k_h,k_h]\times[0,\qmax]$, where $\qmax=2\khmax$ is a cutt-off for large $q$ (i.e. large $k_{1h}$ and $k_{2h}$). We use logarithmic grids also for $p$ and $q$: We note $p = \mp k_h \pm a$ where $a$ takes value on a logarithmic grid of $M_p$ points between $\amin$ and $k_h$ (see equation (\ref{eq:LogGrid})). To save computational time, we take $M_p=\max(8,i_h)$ where $i_h$ is the $k_h$ index on the $(k_h,k_z)$ grid. For $q$, we use $M_q=2 M_h$ points grid between $\qmin$ and $\qmax=2 \khmax$. We choose $\amin=\khmin/M_h$ and $\qmin=\khmin/M_q$ such that increasing resolution allows to improve accuracy at the border of the kinematic box. An illustration of the grid is given in Fig.\ref{figure1}(b). With these choices, the numerical cost is $\sim M_h \times M_z \times M_p \times M_q \sim M_h^3 \times M_z$ operations per time step. \\

	\section{Integration and interpolation schemes}
	
	Integration on logarithmic grids are performed by using the trapezoidal rule after a change of variable. For example, integration in $q$ gives
	\begin{equation}
		\int\limits_{q_i}^{q_{i+1}} ~ F(p,q) ~ \diff q = \int\limits_{i}^{i+1} ~ F(p,q) ~ q \ln \lambda_q ~ \diff i \simeq \frac{F(p,q_i) q_i+ F(p,q_{i+1}) q_{i+1}}{2} ~ \ln \lambda_q.
	\end{equation}
	For $q \leq \qmin \ll \khmin$, we have an integrable singularity with $\Delta(p,q) \simeq \frac{1}{2} \sqrt{2 (k_h^2 - p^2) \delta k_h^2}$ where $\delta = q/k_h \ll 1$, so the integration in $q$ gives
	\begin{equation}
		\int\limits_{0}^{\qmin} ~ \frac{L(p,q)}{\Delta(p,q)} ~ \diff q \simeq \left( \frac{L(p,0) + L(p,\qmin)}{\sqrt{k_h^2 - p^2}} \right) ~ \sqrt{\frac{2\qmin}{k_h}},
	\end{equation}
	where the standard trapezoidal rule (without change of variable) is used and $L(p,0)$ is given by the maximum between $0$ and a linear extrapolation. The contribution of the region $q \geq \qmax$ is not computed and is negligible because of dissipation. The integration scheme is easily adapted to the variable $p$, and generalized to 2D. We checked that this integration scheme allows us to reach order 2 precision. In order to compute $n_1=(k_{1h},|k_{1z}|)$ and $n_2=(k_{2h},|k_{2z}|)$, we interpolate with a bilinear fit of the form $\nk = c_0 + c_h k_h + c_z |k_z| + \beta k_h |k_z|$, where $(c_0,c_h,c_z,\beta)$ are constants in each cell. This interpolation scheme is of order 2 precision. For extrapolations, we simply fix $\nk$ to the maximum between $0$ and a linear fit whose coefficients are fixed with the two first grid points. We checked that the collision integral conserves well the energy for some test spectra. Namely, the energy conservation ratio $\int ~ \ok \Stk ~ \diff \kk / \int ~ \ok |\Stk| ~ \diff \kk$ (see \citet{eden_numerical_2019}) remains less than few $\%$ in our simulations, and decreases as $M^{-2}$ for $M=M_h=M_z$.

	\section{Time stepping}
	
	For the time evolution, we use the splitting method \cite{mclachlan_quispel_splitting_2002}. Namely, the dissipative operator is treated implicitly for half a time step. Then, we use the Runge-Kutta 2 method to treat the collision integral and forcing terms. Finally, we apply again the dissipation operator for half a time step. This method allows us to achieve second-order precision. We employ adaptative time-stepping to follow rapid changes in the wave-action spectrum, and to save computational time when slow changes occur. For this, we compute the ratio between the time step $\dt$ and the minimum of the nonlinear time $\tnl = |\nk / \Stk|$ (which is the inverse of the Boltzmann rate used in other studies, see e.g. \cite{lvov_resonant_2012}) every time step. If $\dt/ \min \limits_{\kk} \tnl > 0.5$, we decrease the time step by a factor $1.25$. Conversely, if $\dt/ \min \limits_{\kk} \tnl < 0.05$, we increase the time step by a factor $1.25$.

	\section{Effect of random initial condition perturbations}
	
	To check the influence of initial perturbations, we have run other simulations employing the log-normal forcing with $\kfh=\kfz=0.07$, but with an initial perturbation for the wave-action spectrum:
	\begin{equation}
		\label{eq:random}
		\nk(t=0) = \frac{a |\eta_{\kk}|}{\ok} ,
	\end{equation}
	where $a$ is the noise amplitude and $\eta_{\kk}$ is a random variable distributed according to the normal law. It corresponds to a statistically homogeneous perturbation of the 3D energy spectrum $\nk \ok$. For these simulations, we used $M_h = M_z = 40$ grid points. In Fig.\ref{fig:random}, we show the 2D energy spectrum at different times for two simulations with $a=0.01$ and $a=1$. We observe that the same interactions (local and nonlocal) are responsible for the evolution of the energy spectrum. Yet, when the noise amplitude is larger, it takes a longer time to see the resonance lines emerge from the background noise (Fig.\ref{fig:random}(d)) when compared to a simulation with a small noise amplitude (Fig.\ref{fig:tophat}(b)). Interestingly, we see that the spectrum is smoothed out in regions near the nonlocal interactions branches toward large wave vectors, shown by dashed lines in Fig.\ref{fig:random}(c) and (f). It shows that the forcing can interact directly with small scales through non-local interactions. Yet, it does not significantly change the evolution of the internal gravity wave spectrum, at least for this set of parameters.
	\begin{figure}[h]
		\centering
		\includegraphics[width=\linewidth]{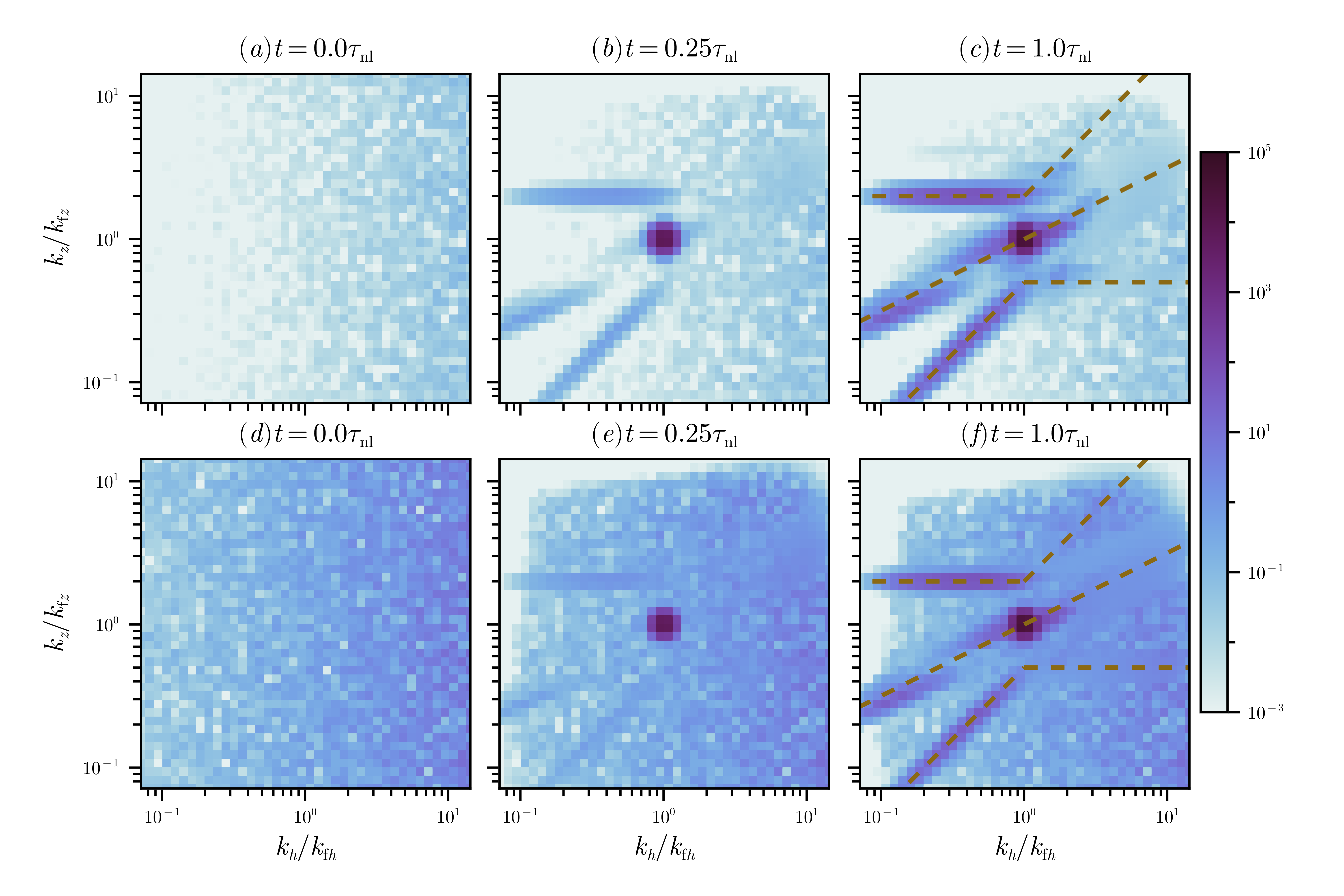}
		\caption{2D energy spectrum $e(k_h,k_z,t)$ at different times for two simulations with two different initial noise amplitude (\ref{eq:random}): (a-c) Small noise amplitude $a=0.01$ and (d-f) Large noise amplitude $a=1$. Dashed lines are for non local interactions (see Fig.2 of the manuscript). \label{fig:random}}
	\end{figure}
	
	\section{Effect of forcing shape and width}
	
	To check the influence of the forcing properties, we have run other simulations employing a top-hat forcing:
	\begin{equation}
		\label{eq:tophat_forcing}
		\fk = \frac{f_0}{\ok} \Theta (k_h - \kfh/w) \Theta (k_h + \kfh w) \Theta (k_z - \kfz/w) \Theta (k_z + \kfz w),
	\end{equation}
	where $\Theta$ is the Heaviside function, $\kfh = \kfz = 0.07$ is the forcing position, $w$ the forcing width factor, and $f_0$ is such that the total energy injection rate is equal to unity. For these simulations, we used $M_h = M_z = 40$ grid points. In Fig.\ref{fig:tophat}, we show the energy spectrum at different times for two simulations with $w=1.25$ and $w=2$. We observe that the same interactions (local and nonlocal) are responsible for the evolution of the energy spectrum. Yet, a larger forcing width involves more resonances, spreading energy faster at early times (Fig.\ref{fig:tophat}(d)), when compared to a simulation with a narrow forcing (Fig.\ref{fig:tophat}(b)). In particular, the direct energy transfer (to large wave vectors) is faster with a wide forcing range. Interestingly, the last state of the energy spectrum is very similar for $w=1.25$ (Fig.\ref{fig:tophat}(c)) and $w=2$ (Fig.\ref{fig:tophat}(f)).
	\begin{figure}[h]
		\centering
		\includegraphics[width=\linewidth]{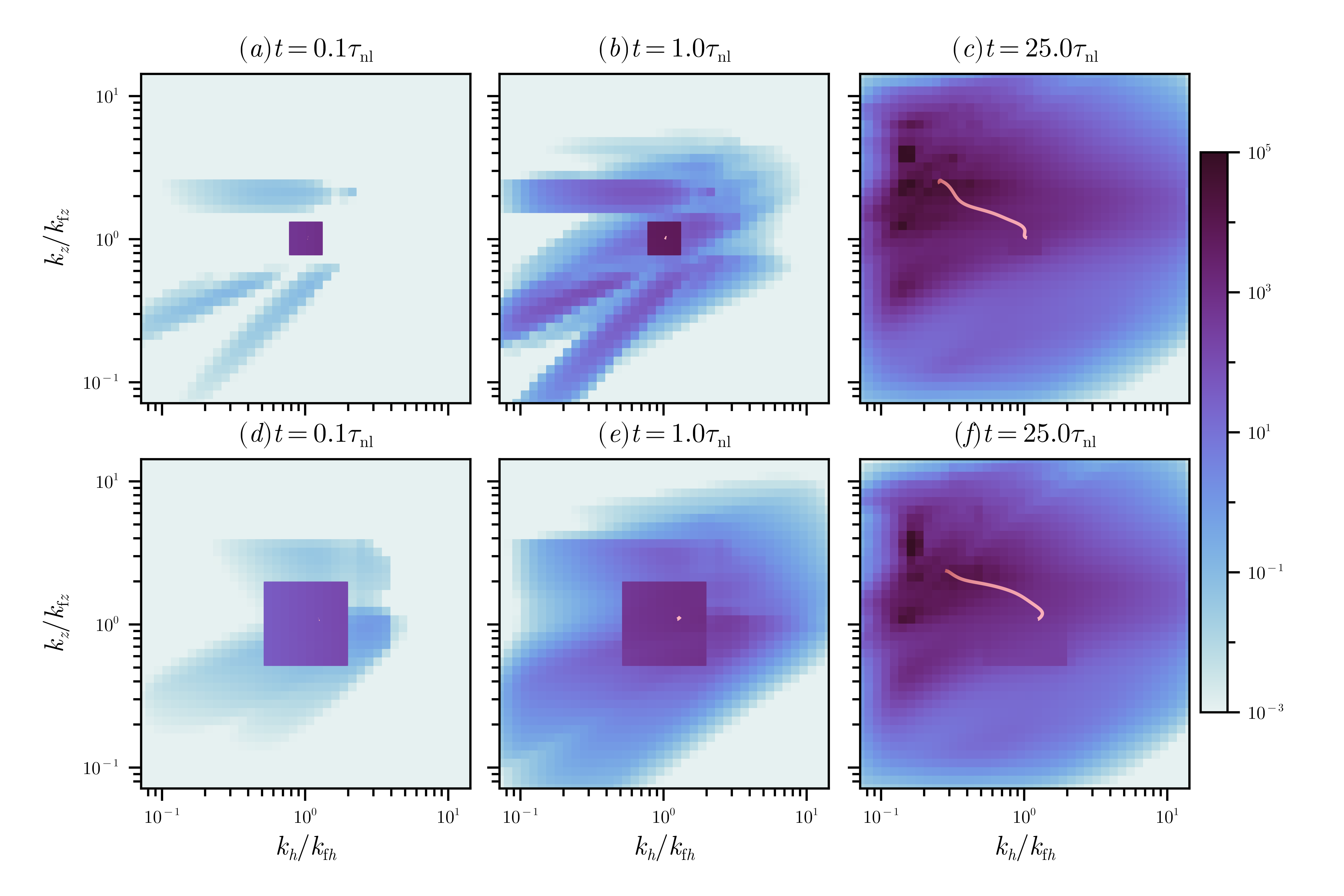}
		\caption{2D energy spectrum $e(k_h,k_z,t)$ at different times for two simulations with top-hat forcing with different width (\ref{eq:tophat_forcing}): (a-c) Small forcing width $w=1.25$ and (d-f) Large forcing width $w=2$. The colored lines shows the trajectory of the integral scales (see Fig.4 of the manuscript). \label{fig:tophat}}
	\end{figure}

\end{widetext}

\end{document}